\documentclass[showpacs,doublespace,preprintnumbers,amsmath,amssymb,pre]{revtex4}
\usepackage{bm}
\usepackage{graphicx}

\begin{document}
\begin{large}

\title{Percolation of aligned rigid rods on two-dimensional triangular lattices}

\author{P. Longone}
\affiliation{Departamento de F\'{\i}sica, Instituto de F\'{\i}sica
Aplicada, Universidad Nacional de San Luis-CONICET, Chacabuco 917,
D5700BWS San Luis, Argentina}
\author{P. M. Centres}
\affiliation{Departamento de F\'{\i}sica, Instituto de F\'{\i}sica
Aplicada, Universidad Nacional de San Luis-CONICET, Chacabuco 917,
D5700BWS San Luis, Argentina}
\author{A.J. Ramirez-Pastor$^{\dag}$}
\affiliation{Departamento de F\'{\i}sica, Instituto de F\'{\i}sica
Aplicada, Universidad Nacional de San Luis-CONICET, Chacabuco 917,
D5700BWS San Luis, Argentina}

\date{\today}

\begin{abstract}
The percolation behavior of aligned rigid rods of length $k$ ($k$-mers) on two-dimensional triangular lattices has been studied by numerical simulations and finite-size scaling analysis. The $k$-mers, containing $k$ identical units (each one occupying a lattice site), were irreversibly deposited along one of the directions of the lattice. The connectivity analysis was carried out by following the probability $R_{L,k}(p)$ that a lattice composed of $L \times L$ sites percolates at a concentration $p$ of sites occupied by particles of size $k$. The results, obtained for $k$ ranging from 2 to 80, showed that the percolation threshold $p_c(k)$ exhibits a increasing function when it is plotted as a function of the $k$-mer size. The dependence of $p_c(k)$ was determined, being $p_c(k)=A+B/(C+\sqrt{k})$, where $A = p_c(k \rightarrow \infty)= 0.582(9)$ is the value of the percolation threshold by infinitely long $k$-mers, $B =-0.47(0.21)$ and $C = 5.79(2.18)$. This behavior is completely different to that observed for square lattices, where the percolation threshold decreases with $k$. In addition, the effect of the anisotropy on the properties of the percolating phase was investigated. The results revealed that, while for finite systems the anisotropy of the deposited layer favors the percolation along the parallel direction to the nematic axis, in the thermodynamic limit, the value of the percolation threshold is the same in both parallel and transversal directions. Finally, an exhaustive study of critical exponents and universality was carried out, showing that the phase transition occurring in the system belongs to the standard random percolation universality class regardless of the value of $k$ considered.
\end{abstract}


\maketitle

\noindent $\dag$ To whom all correspondence should be addressed.

\newpage

 \section{Introduction} \label{intro}

Percolation is a very active field of research and applied to a wide range of fields, such as biology, nanotechnology, device physics, physical chemistry, and materials science \cite{Stauffer,Sahimi,Grimmett,Bollo}. The problem of percolation is not a new one but still attracts considerable interest \cite{Kundu,Kundu1,Treffen}, and some unsolved questions remain.

Percolation theory was derived for periodic lattices of sites (bonds) which are occupied with probability $p$ or empty (nonoccupied) with probability $(1 - p)$ \cite{Stauffer}. In the case of deposition processes, $p$ coincides (in the thermodynamic limit) with the coverage of the lattice or fraction of sites occupied by the deposited objects. If the concentration of these objects is sufficiently large, a cluster (a group of occupied sites in such a way that each site has at least one occupied nearest-neighbor site) extends from one side to the opposite one of the system. The central idea of the percolation theory is based in finding the minimum concentration for which a complete path of adjacent sites crossing the entire system becomes possible. This value of the concentration rate is named the critical concentration or percolation threshold $p_c$, and determines the phase transition in the system \cite{Stauffer}.

One of the most popular methods of studying percolation of deposited objects is the Random Sequential Adsorption (RSA) technique \cite{Feder,Evans,Talbot}. In this process, objects of a specified shape are randomly and sequentially adsorbed onto a substrate and then immobilized. Excluded volume, or particle-particle interaction, is incorporated by rejection of deposition overlap, while particle-substrate interaction is modeled by the irreversibility of deposition. The final state generated by RSA is a disordered state (known as jamming state), in which no more objects can be deposited due to the absence of free space of appropriate size and shape (the jamming state has infinite memory of the process and the orientational order is purely local). Thus, a competition between percolation and jamming is established \cite{Evans,Talbot}. In some applications one may want that percolation dominates (i.e. communications) in others one may prefer that jamming dominates and percolation is suppressed at an early stage (i.e. forest fires).

For randomly distributed and isotropically oriented linear $k$-mers \footnote{Also denoted as needles, rods, or sticks.} (linear rigid particles occupying $k$ consecutive sites) on square lattices, it was shown that the percolation threshold does not change monotonically with the length of needles \cite{Tara2012,Tara2015,JSTAT3}. For short objects the percolation threshold decreases rapidly, goes through a minimum around $k = 13 \dots 15$, and then it started to increase moderately. Later, Kondrat {\it et al.} \cite{Kondrat2017} presented a strict proof that in any jammed configuration of nonoverlapping, fixed-length, horizontal or vertical needles on a square lattice, all clusters are percolating clusters. The theorem refutes the conjecture \cite{Tara2012,Tara2015,JSTAT3} that in the RSA of such needles on a square lattice, percolation does not occur if the needles are longer than some threshold value $k^*$, estimated to be of the order of several thousand.

In a very recent paper, Slutskii {\it et al.} \cite{Tara2018}, using simulation techniques, corroborated the result reported by Kondrat {\it et al.} \cite{Kondrat2017}. Based in a
very efficient parallel algorithm, the authors studied the problem of large linear $k$-mers (up to $k=2^{17}$) on a square lattice with periodic boundary conditions. The obtained results indicate that the percolation threshold tends to a constant value as $k \rightarrow \infty$, being $p_{c}(k \rightarrow \infty)=0.615(1)$. The limit value of $p_{c}$ is lower than the asymptotic value of the jamming coverage: $p^{j}(k \rightarrow \infty)=0.655(9)$ \cite{Lebo2011}. This finding reinforces the theoretical analysis in Ref. \cite{Kondrat2017}, namely, in the case of linear $k$-mers on square lattices, percolation always occurs before jamming.

An interesting problem arises when the probability of taking horizontal and vertical orientation is not the same. In this context, the advent of modern techniques for building highly conductive rodlike particles (such as carbon nanotubes \cite{Charlier}, metal nanowires \cite{Sofiah}, etc.) has considerably encouraged the investigation of anisotropic composites made of these elongated particles on an insulating matrix. The study of the conductive properties of these composite materials is an area of increasing interest for the production of flexible transparent conductors \cite{Hecht,McCoul,Mutiso}, with diverse applications in solar cells, touch-screens, and transparent heaters \cite{De,Mutiso1,Ackermann,Kumar1,Kumar2}. These promising applications are inspiring both theoretical and experimental studies in this field \cite{Taherian}.

In order to design a composite with the desired properties, it is crucial to understand and control the formation of a system-spanning network of nanofillers inside the host matrix, which happens above a critical concentration of filler material. This critical concentration coincides with the percolation threshold of the system \cite{Finner2018,Torquato}, demonstrating the importance of percolation theory and its applicability to study the electrical conductivity of materials composed of rodlike highly conducting fillers. In this line, numerous works have been conducted on percolation of rodlike particles and its connection with the electrical conductivity \cite{Pike,Balberg1,Balberg2,Balberg3,Chatter,Tara2016,Lebo2018,Tara2018a}. The studies in Refs. \cite{Pike,Balberg1,Balberg2,Balberg3,Chatter,Tara2016,Lebo2018,Tara2018a} represent an important step in the understanding of the percolating properties of anisotropic conductors.

In a previous paper from our group, the effect of anisotropy (or $k$-mer alignment) on percolation was investigated for the case of aligned rigid $k$-mers on square lattices \cite{PRE7}. The results, obtained for $k$ ranging from 1 to 14, showed that $(i)$ the percolation threshold exhibits a decreasing function when it is plotted as a function of the $k$-mer size; and $(ii)$ for
any value of $k (k > 1)$, the percolation threshold is higher for aligned rods than for rods isotropically deposited. Later, Tarasevich {\it et al.} \cite{Tara2012} extend the analysis in Ref. \cite{PRE7} to larger lattices ($100 \leq L \leq 19200$) and longer objects ($2 \leq k \leq 512$). The authors corroborate the results obtained by Longone {\it et al.} \cite{PRE7} for the case of perfectly aligned rods, and complete the study by including the percolation behavior of partially ordered phases (states whose degree of anisotropy varies between the two limit cases, i.e. isotropic and perfectly aligned $k$-mers).

In the case of triangular lattices, many interesting results have been reported on RSA of objects of various shape \cite{Budi1997}, reversible RSA \cite{Budi2005}, reversible RSA of mixtures \cite{Lonca2007}, anisotropic RSA of extended objects \cite{Budi2011}, percolation of extended objects \cite{Budi2012} and jamming and percolation in RSA of extended objects on lattice with quenched impurities \cite{Budi2016}. However, the effect of $k$-mer alignment on percolation has been poorly studied. In this context, the main objective of the present paper is to study the percolation behavior of aligned rigid rods on 2D triangular lattices. For this purpose, extensive numerical simulations (with $2 \leq k \leq 80$ and
$75 \leq L/k \leq 640$) supplemented by analysis using finite-size scaling theory have been carried out. The obtained results revealed that the percolation threshold $p_c(k)$ is an increasing function with $k$. This finding contrasts with the decreasing tendency observed for $p_c(k)$ in square lattices, showing that (1) it is of interest and of value to inquire how a specific lattice structure influences the main percolation properties of particles occupying more than one site; and (2) the structure of the lattice plays a fundamental role in determining the statistics of extended objects. In addition, the anisotropy effect on the percolation probabilities characterizing the different lattice directions was investigated. The study also includes a complete analysis of critical exponents and universality.

The present work is a natural extension of our previous research in the area of percolation of polyatomic species and the results obtained here could have potential application in the field of conductivity in composite materials. The paper is organized as it follows: the model and basic definitions are given in Sec. \ref{model}. Percolation properties are studied in Sec. \ref{perco}. The conclusions are drawn in Sec. \ref{conclu}. Finally, a complete study of critical exponents and universality is presented in the Supplemental Material \cite{SM}.

 \section{Model and basic definitions} \label{model}

Straight rigid rods are deposited randomly, sequentially and irreversibly on a 2D triangular lattice. In the computer simulations, a rhombus-shaped system of $M = L \times L$ sites ($L$ rows and $L$ columns) is used (see Fig. \ref{Fig1}). The deposition process is performed with the following restrictions: $(1)$ the $k$-mers contain $k$ identical units and each one occupies a lattice site. Small adsorbates with spherical symmetry would correspond to the monomer limit ($k=1$); $(2)$ the distance between $k$-mer units is assumed in registry with the lattice constant $a$; hence exactly $k$ sites are occupied by a $k$-mer when deposited; $(3)$ the $k$-mers are deposited along one of the directions of the lattice, forming a nematic phase as depicted in Fig. \ref{Fig1}; $(4)$ the incoming particles must not overlap with previously added objects; and $(5)$ periodic boundary conditions are considered.

\begin{figure}
\begin{center}
\includegraphics[width=0.95\columnwidth]{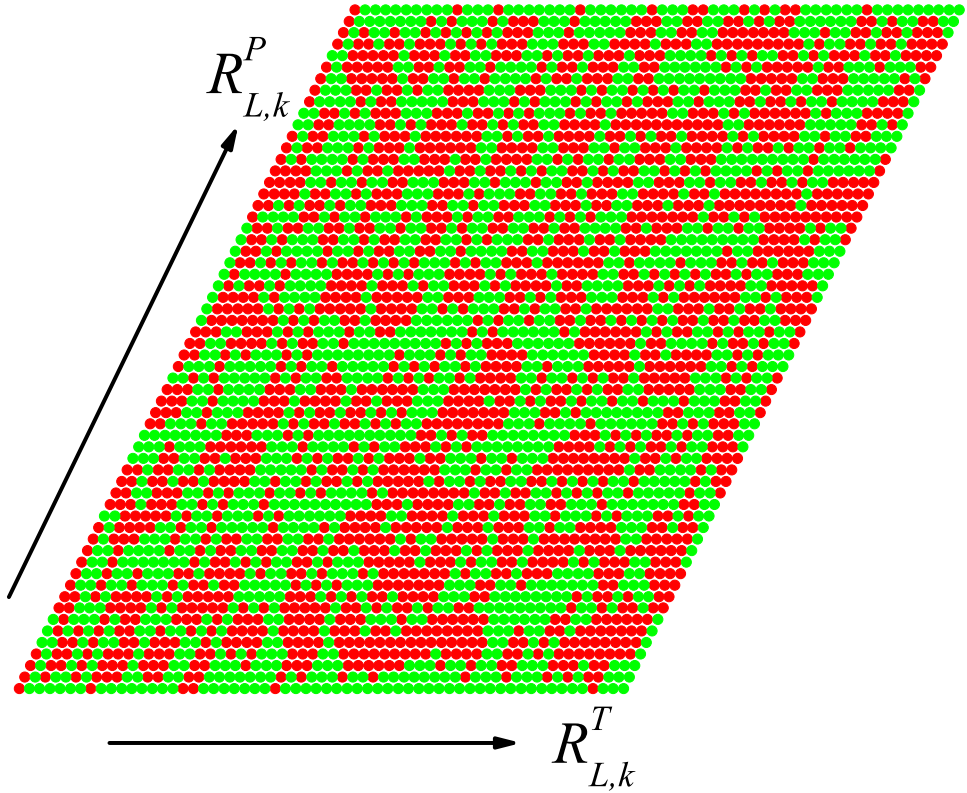}
\caption{Snapshot corresponding to a configuration of aligned tetramers ($k=4$) on a rhombus-shaped triangular lattice. Green and red circles represent empty sites and tetramer units, respectively.}
\label{Fig1}
\end{center}
\end{figure}

Due to the blocking of the lattice by the already randomly adsorbed elements, the limiting or jamming coverage, $p^j=p(t=\infty)$ is less than that corresponding to the close
packing ($p^j<1$). Note that $p(t)$ represents the fraction of lattice sites covered at time $t$ by the deposited objects. Consequently, $p$ ranges from $0$ to $p^j$ for objects occupying more than one site \cite{Evans}. For a fully aligned system, as studied here, the jamming problem reduces to the one-dimensional (1D) case. In this limit, $p(t)$ can be written as \cite{Lebo2011,Redner},
\begin{equation}
p(t) = k \int_0^t \exp \left[ -u-2 \sum_{j=1}^{k-1} \left( \frac{1-e^{-ju}}{j} \right) \right]du. \label{pt}
\end{equation}
The numerical evaluation of Eq. (\ref{pt}) allows us to obtain the dependence on $k$ of the jamming coverage $p^j(k)$. Note that Eq. (\ref{pt}) corresponds to an exact result obtained for an infinite 1D lattice.

\section{The percolation threshold} \label{perco}

As mentioned in Sec. \ref{intro}, the central idea of the pure percolation theory is based in finding the minimum concentration of elements (sites or bonds) for which a cluster extends from one side to the opposite one of the system. For this particular value of the concentration rate, the percolation threshold $p_c$, at least one spanning cluster (also called ``infinite" cluster, inspired by the thermodynamic limit) connects the borders of the system \cite{aizen,cardy,shchur1,shchur2}. In that case, a second-order phase transition appears at $p_c$ which is characterized by well defined critical exponents.

In the simulations, each run consists of the following stages: (a) the construction of the lattice for the desired fraction $p=kN/M$ of sites ($N$ is the number of $k$-mers deposited), according to the filling procedure presented in previous section; and (b) the cluster analysis by using the Hoshen and Kopelman algorithm \cite{Hoshen1,Hoshen2} with open boundary conditions. In the last step, the size of the largest cluster $S_L$ is determined, as well as the existence of a percolating island. For this purpose, the probability $R=R^X_{L,k}(p)$ that a $L \times L$ lattice percolates at a concentration $p$ of sites occupied by rods of size $k$ can be defined. Here, the following definitions can be given according to the meaning of $X$ \cite{Stauffer,Yone1,Yone2}:
\begin{itemize}
  \item $R^{P} _{L,k}(p)$: the probability of finding a percolating cluster in a parallel direction to the nematic alignment (see Fig. \ref{Fig1}),\\
  \item $R^{T} _{L,k}(p)$: the probability of finding a percolating cluster in a transverse direction to the nematic alignment (see Fig. \ref{Fig1}).
  \end{itemize}
Other useful definitions for the finite-size analysis are:
\begin{itemize}
  \item $R^{U}_{L,k}(p)$: the probability of finding either a parallel or a transverse percolating cluster,\\
  \item $R^{I}_{L,k}(p)$: the probability of finding a cluster which percolates both in a parallel and in a transverse direction,\\
  \item $R^{A}_{L,k}(p)=\frac{1}{2}[R^{U}_{L,k}(p)+R^{I}_{L,k}(p)]$.
\end{itemize}
$n$ runs of such two steps are carried out for obtaining the number $m^X$ of them for which a percolating cluster of the desired criterion $X$ is found. Then, $R^{X}_{L,k}(p)=m^X/n$ is defined and the procedure is repeated for different values of $p$, $L$ and $k$. A set of $n= 10^6$ independent samples is numerically prepared for each pair $p$ and $L/k$ ($L/k=75$, $100$, $128$, $256$ and $640$). The $L/k$ ratio is kept constant to prevent spurious effects due to the $k$-mer size in comparison with the lattice linear size $L$.

\begin{figure}
\begin{center}
\includegraphics[width=0.60\columnwidth]{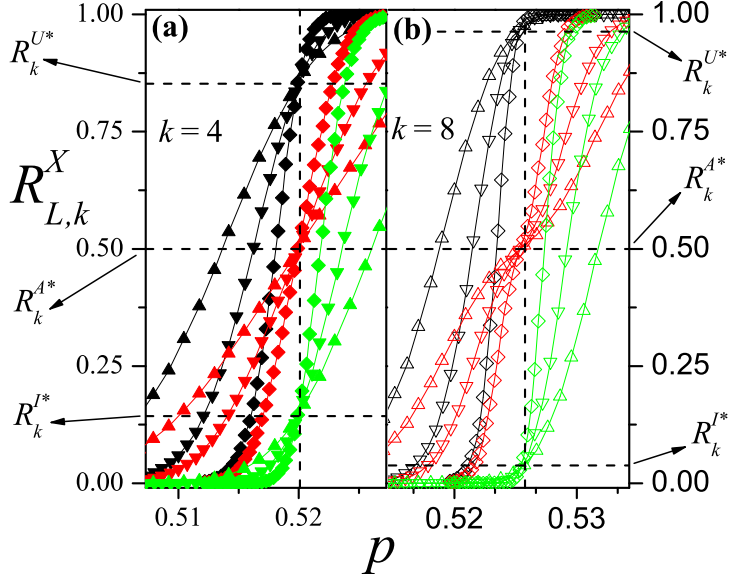}
\caption{Fraction of percolating lattices $R^X_{L,k}(p)$ ($X   =   I, U, A$ as indicated) as a function of the concentration $p$ for $k$ = 4 (a), $k=8$ (b) and three different lattice sizes: $L/k = 128$ (up triangles), $L/k = 256$ (down triangles) and $L/k = 640$ (diamonds). In each panel, the vertical dashed line denotes the percolation threshold in the thermodynamic limit.}
\label{FigR}
\end{center}
\end{figure}

In Fig. \ref{FigR}, the probabilities $R^{A}_{L,k}(p)$, $R^{I}_{L,k}(p)$ and $R^{U}_{L,k}(p)$ are presented for aligned rods with $k=4$ (Fig. \ref{FigR}a), $k = 8$ (Fig. \ref{FigR}b). As mentioned in the previous paragraph, the simulations were performed for lattice sizes ranging between $L/k = 75$ and $L/k = 640$. For clarity, three sizes are shown in the figure: $L/k = 128$ (up triangles), $L/k = 256$ (down triangles) and $L/k = 640$ (diamonds). Several conclusions can be drawn from Fig. \ref{FigR}. First, curves for different lattice sizes but with the same value of $k$ cross each other in a unique point, $R^{X^*}_k$ (measured in the vertical axis, see figure), which depends on the criterion $X$ used and those points are located at very well defined values in the $p$-axes determining the critical percolation threshold (measured in the horizontal axis, see figure) for each $k$. Second, $p_c(k)$ shifts to the left upon increasing the $k$-mer size. This observation is a clear indication of that the percolation threshold decreases upon increasing $k$.

\begin{figure}
\begin{center}
\includegraphics[width=0.60\columnwidth]{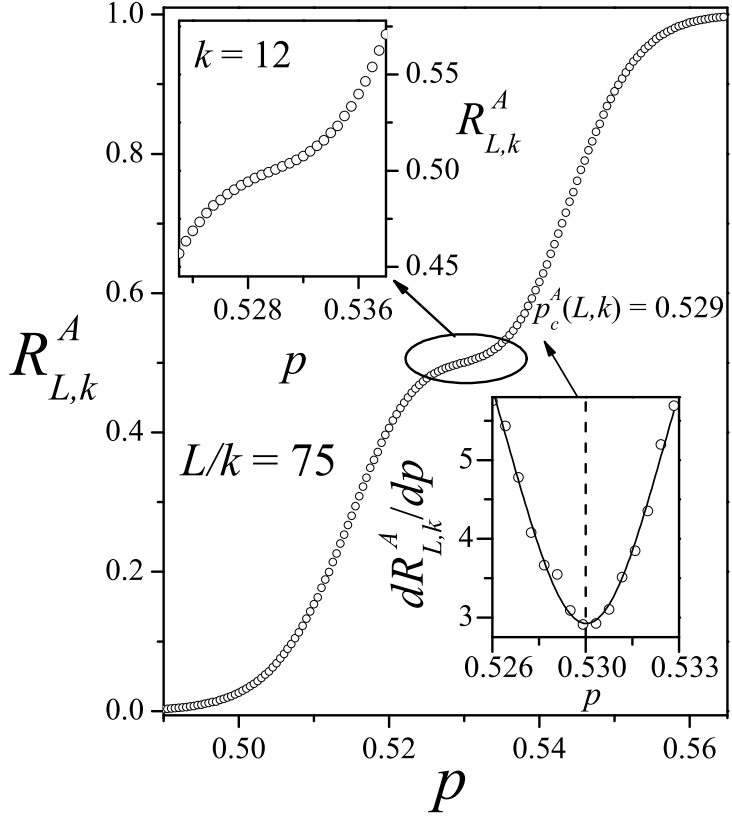}
\caption{Percolation probability $R^{A}_{L,k}(p)$ as a function of the concentration $p$ for $k = 12$ and $L/k=75$. Upper-left inset: Magnification of the main figure in the range $0.523\leq p \leq 0.538$. Lower-right inset: $dR^{A}_{L,k}/dp$ as a function of $p$ around the critical point $p_{c}^{A}(L,k)$. Symbols correspond to simulation data and the solid line represents a Gaussian fitting curve.}
\label{minRA}
\end{center}
\end{figure}

Third, $R^{I^*}_k$ and  $R^{U^*}_k$ show a strong dependence on the $k$-mer size. For $k=1$ (data not shown here), $R^{I^*}_1 \approx 0.311$ and $R^{U^*}_1 \approx 0.687$, as reported in previous work for standard site percolation on a rhombus-shaped lattice with open boundary conditions \cite{JSTAT5}. As $k$ is increased, two well-differentiated behavior are observed: (1) $R^{U^*}_k$ increases monotonically to $R^{U^*}_k \approx 1 $ for larger sizes; and (2) $R^{I^*}_k$ decreases monotonically to $R^{I^*}_k \approx 0 $ for larger sizes. On the other hand, $R^{A^*}_k$ remains constant (around 0.5 \cite{JSTAT5}) when $k$ increases. A similar behavior has been observed in the case of aligned $k$-mers on square lattices \cite{PRE7} and thermal transitions in the presence of anisotropy \cite{Selke1,Selke2}.

In percolation theory, the value of the probability $R^{X}_{L}$ at the transition point in the thermodynamic limit plays an important role in the scaling theory, being indicative of the universality class of the transition. From this perspective, the dependence of $R^{I^*}_k$ and $R^{U^*}_k$ on $k$ could be taken as a first indication of a nonuniversal behavior of the system for variable $k$-mer size. However, as pointed out by Selke et al. \cite{Selke1,Selke2}, the measure of the probability intersection may depend on various details of the model which do not affect the universality class, in particular, the boundary condition, the shape of the lattice, and the anisotropy of the system. Consequently, more research is required to determine the universality class of the phase transition.

In order to express $R^{X}_{L,k}(p)$ as a function of continuous values of $p$, it is convenient to fit $R^{X}_{L,k}(p)$ with some approximating function through the least-squares method. The fitting curve is the {\em error function} because $dR^{X}_{L,k}/dp$ is expected to behave like the Gaussian distribution \footnote{Even though the behavior of $dR^{X}_{L,k}/dp$ is known not to be a gaussian in all range of $p$, this quantity is approximately gaussian near the peak, and Eq. (\ref{ecu1}) is a good approximation for the purpose of locating its maximum \cite{Newman}.}
\begin{equation}\label{ecu1}
    \frac{dR^{X}_{L,k}}{dp}=\frac{1}{\sqrt{2\pi}\Delta^{X}_{L,k}}\exp \left\{ -\frac{1}{2} \left[\frac{p-p_{c}^{X}(L,k)}{\Delta^{X}_{L,k}}
    \right]^2 \right\},
\end{equation}
where $p_{c}^{X}(L,k)$ is the concentration at which the slope of $R^{X}_{L,k}(p)$ is the largest and $\Delta^{X}_{L,k}$ is the standard deviation from $p_{c}^{X}(L,k)$.

The standard procedure described in the last paragraph is valid for $R^{T}_{L,k}(p)$ and $R^{P}_{L,k}(p)$ in all range of $k$. The same does not occur in the case of $R^{A}_{L,k}(p)$. In fact, as will be discussed in detail later (see Figs. \ref{FigRTP} and \ref{Figdist}), the anisotropy of the percolating phase leads to a separation between the parallel and transversal probabilities. As a consequence of this separation, which increases with $k$, the $R^{A}_{L,k}(p)$ curves tend to gradually develop a plateau, with a marked inflection point around $R^{A^*}_k \approx 0.5$. This singularity, which is barely perceptible in Fig. \ref{FigR}, can be clearly visualized in Fig. \ref{minRA}, where the percolation probability $R^{A}_{L,k}(p)$ has been plotted as a function of the concentration $p$ for $k = 12$ and $L/k=75$.

The upper-left inset shows a zoom of the plateau region. On the other hand, the lower-right inset shows $dR^{A}_{L,k}/dp$ as a function of $p$ around the inflection point. Thus, the value of $p_{c}^{A}(L,k)$ can be obtained from the concentration at which the minimum occurs. For an accurate determination of this concentration, we fit the simulation data with an inverted Gaussian function. The procedure is shown in the lower-right inset: open circles correspond to simulation data and solid line represents the Gaussian fitting curve.

Once determined the positions $p_{c}^{X}(L,k)$, the percolation threshold $p_{c}(k)$ can be obtained using an extrapolation scheme. Thus, for each criterion ($I$, $U$ and $A$), and for each value of $k$, one expects that \cite{Stauffer}
\begin{equation}
p_{c}^{X}(L,k)= p_{c}(k) + A^X L^{-1/\nu},
\label{extrapolation}
\end{equation}
where $A^X$ is a nonuniversal constant and $\nu$ is the critical exponent of the correlation length which will be taken as 4/3 for the present analysis, since, as it will be shown in the Supplemental Material \cite{SM}, our model belongs to the same universality class as 2D random percolation \cite{Stauffer}.

\begin{figure}
\begin{center}
\includegraphics[width=0.60\columnwidth]{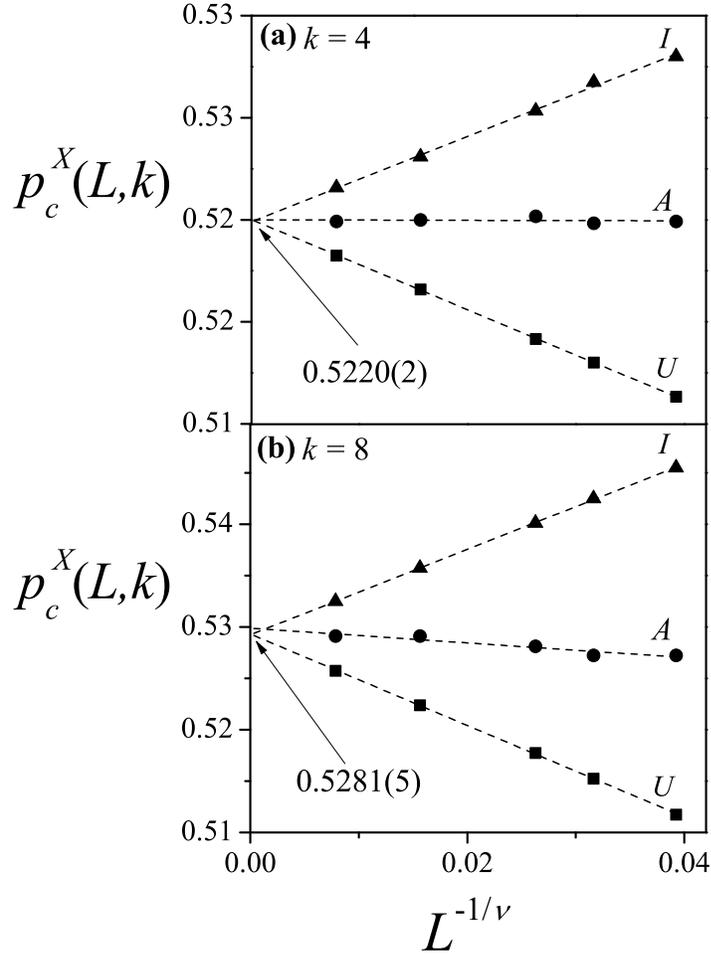}
\caption{Extrapolation of $p_{c}^{X}(L,k)$ towards the thermodynamic limit according to the theoretical prediction given by Eq. (\ref{extrapolation}). Triangles, circles and squares  denote the values of $p_{c}^{X}(L,k)$ obtained by using the criteria $I$, $A$ and $U$, respectively. The data correspond to the cases presented in Fig. \ref{FigR}: $k$ = 4 (a) and $k=8$ (b).}
\label{extrapo}
\end{center}
\end{figure}

Figure \ref{extrapo} shows the plots towards the thermodynamic limit of $p_{c}^{X}(L,k)$ according to Eq. (\ref{extrapolation}) for the data in Fig. \ref{FigR}. From extrapolations it is possible to obtain the percolation thresholds for the criteria $I$, $A$ and $U$. Combining the three estimates for each case, the final values of $p_{c}(k)$ can be obtained. Additionally, the maximum of the differences between $|p_{c}(k)^{U}-p_{c}(k)^{A}|$ and $|p_{c}(k)^{I}-p_{c}(k)^{A}|$ gives the error bar for each determination of $p_{c}(k)$. In this case, the values obtained were: $p_{c}(k=4)=0.5220(2)$ (a), and $p_{c}(k=8)=0.5281(5)$ (b).

The procedure in Fig. \ref{extrapo} was repeated for different values of $k$ ranging between 2 and 80. The obtained values of $p_c(k)$ are collected in Table I (second column) and are plotted in Fig. \ref{pcvsk} (open squares). As it can be observed from the figure, the percolation threshold increases upon increasing $k$. The curve rapidly increases for small values of $k$, it flatten out for larger values of $k$, and asymptotically converges towards a definite value as $k \rightarrow \infty$. In the range $2 \leq k \leq 80$, the data of $p_c(k)$ can be fitted with the function proposed in Ref. \cite{Tara2018}: $p_c(k)=A+B/(C+\sqrt{k})$, being $A = p_c(k \rightarrow \infty)= 0.582(9)$ the value of the percolation threshold by infinitely long $k$-mers, $B =-0.47(0.21)$ and $C = 5.79(2.18)$. The adjusted coefficient of determination is $R^2=0.9899$. As observed in previous theoretical \cite{Balberg1,Balberg2,Balberg3,Chatter}, experimental \cite{Du,Grossiord,Deng}, and simulation work \cite{Tara2012,PRE7,Rahatekar,White,Kale}, the percolation threshold is higher for aligned rods than for rods isotropically deposited (see Ref. \cite{JSTAT7}, where the problem of isotropic $k$-mers on triangular lattices has been studied).

\begin{table}
\label{T1}
\begin{center}
\caption{Percolation thresholds versus $k$.}
\begin{tabular}{p{1cm} |p{1.5cm}}
$k$ & $p_c(k)$  \\
\hline
2	& 0.5157(2)  \\
4	& 0.5220(2)  \\
8	& 0.5281(5)  \\
12	& 0.5298(8)  \\
16	& 0.5328(7)  \\
32	& 0.5407(6)  \\
48	& 0.5455(4)  \\
64	& 0.5487(8)  \\
80	& 0.5500(6)  \\
\end{tabular}
\end{center}
\end{table}

The inset of Fig. \ref{pcvsk} shows the behavior of $p_c(k)$ for aligned $k$-mers on square lattices. Solid triangles and open diamonds correspond to data in Refs. \cite{PRE7} and \cite{Tara2012}, respectively. The dashed line represents the fitting curve obtained in Ref. \cite{Tara2012}: $p_c(k)= a_1/k^{\alpha_1}+ p_c(k \rightarrow \infty)$, where $p_c(k \rightarrow \infty)= 0.533(1)$, $a_1= 0.088(3)$ and $\alpha_1= 0.72(4)$. These results are qualitatively different from those obtained for triangular lattices (main figure). Clearly, the structure of the lattice plays a fundamental role in determining the statistics and percolation properties of extended objects.

Figure \ref{pcvsk} also includes the behavior of $p^j(k)$ for aligned $k$-mers (solid circles joined by a solid line). The corresponding numerical values were obtained by solving Eq. (\ref{pt}) (with $t \rightarrow \infty$). The curve of $p^j(k)$ remains above the curve of $p_c(k)$, tending to $p^j(k \rightarrow \infty) \approx  0.7475979202$ in the limit of infinitely long rods \cite{Renyi}. This finding indicates that the RSA model of aligned $k$-mers on triangular lattices presents percolation transition in the whole range of $k$.

\begin{figure}
\begin{center}
\includegraphics[width=0.60\columnwidth]{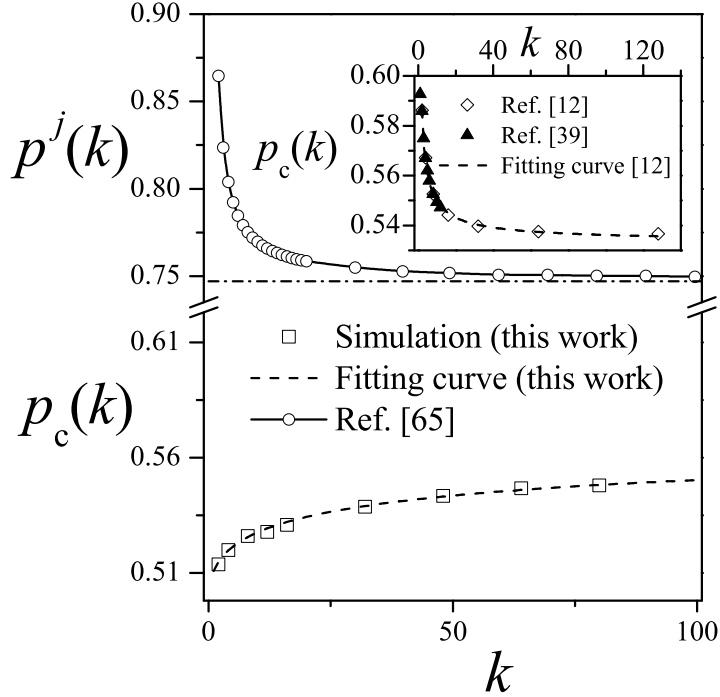}
\caption{Percolation threshold as a function of $k$ for aligned $k$-mers on triangular lattices (open squares). In all cases, the error bar is smaller that the size of the symbols. The dashed line corresponds to the fitting function $p_c(k)=A+B/(C+\sqrt{k})$ (see discussion in the text). The figure also includes the curve of $p^j(k)$ (solid circles joined by a solid line). The corresponding numerical values were obtained by solving Eq. (\ref{pt}) (with $t \rightarrow \infty$). Inset: Percolation threshold as a function of $k$ for aligned $k$-mers on square lattices. Solid triangles and open diamonds denote previous data in Refs. \cite{PRE7} and \cite{Tara2012}, respectively. The dashed line represents the fitting curve obtained in Ref. \cite{Tara2012}: $p_c(k)= a_1/k^{\alpha_1}+ p_c(k \rightarrow \infty)$.}
\label{pcvsk}
\end{center}
\end{figure}

\begin{figure}
\begin{center}
\includegraphics[width=0.55\columnwidth]{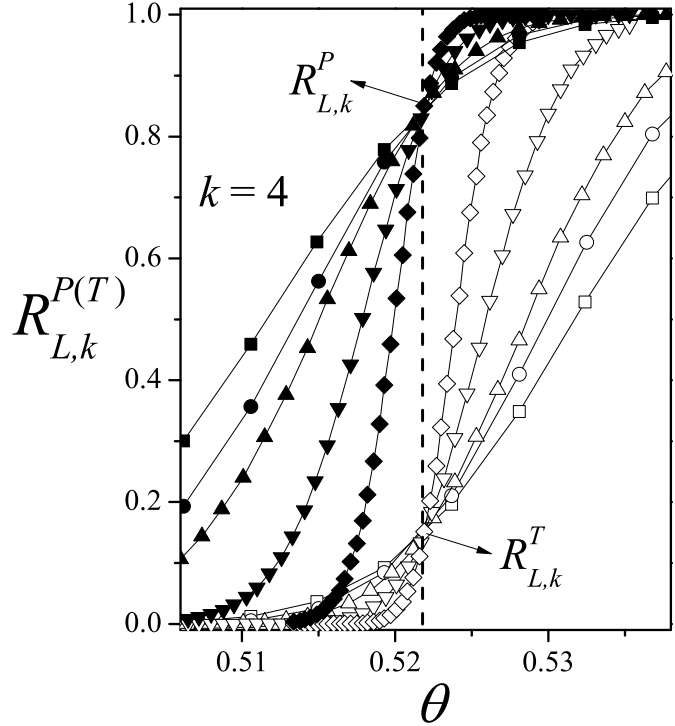}
\caption{Fraction of percolating lattices $R^X_{L,k}(p)$ ($X = P,T$ as indicated) as a function of the concentration $p$ for $k$ = 4 and different lattice sizes: $L/k = 75$ (squares), $L/k = 100$ (circles), $L/k = 128$ (up triangles), $L/k = 256$ (down triangles) and $L/k = 640$ (diamonds). The vertical dashed line denotes the percolation threshold in the thermodynamic limit.}
\label{FigRTP}
\end{center}
\end{figure}

To complete the study, and given the anisotropy of the percolating phase, it is interesting to analyze the behavior of the transversal [$R^{T} _{L,k}(p)$] and parallel [$R^{P} _{L,k}(p)$] percolation probabilities. In Fig. \ref{FigRTP}, the probabilities $R^{P}_{L,k}(p)$ (solid symbols) and $R^{T}_{L,k}(p)$ (open symbols) are presented for a typical case: aligned rods with $k=4$ and different lattice sizes between $L/k=75$ and $L/k=640$. From a simple inspection of the figure it is observed that: $(a)$ for a fixed value of $L=L_1$, the $R^{P}_{L_1,k}(p)$ curve shifts to the left of the $R^{T}_{L_1,k}(p)$ curve. The result indicates that for finite systems the anisotropy of the deposited layer favors the percolation along the direction of the nematic phase. This scenario does not occur in isotropic systems, where for a fixed $L$, the vertical and horizontal percolation probabilities are indistinguishable \cite{Newman}; and $(b)$ $R^{P^*}_k$ and $R^{T^*}_k$ crossing points are located at the same point on the $p$-axis (vertical line in the figure), indicating that, in the thermodynamic limit, the value of the percolation threshold is the same in both parallel and transversal directions.

\begin{figure}
\begin{center}
\includegraphics[width=0.50\columnwidth]{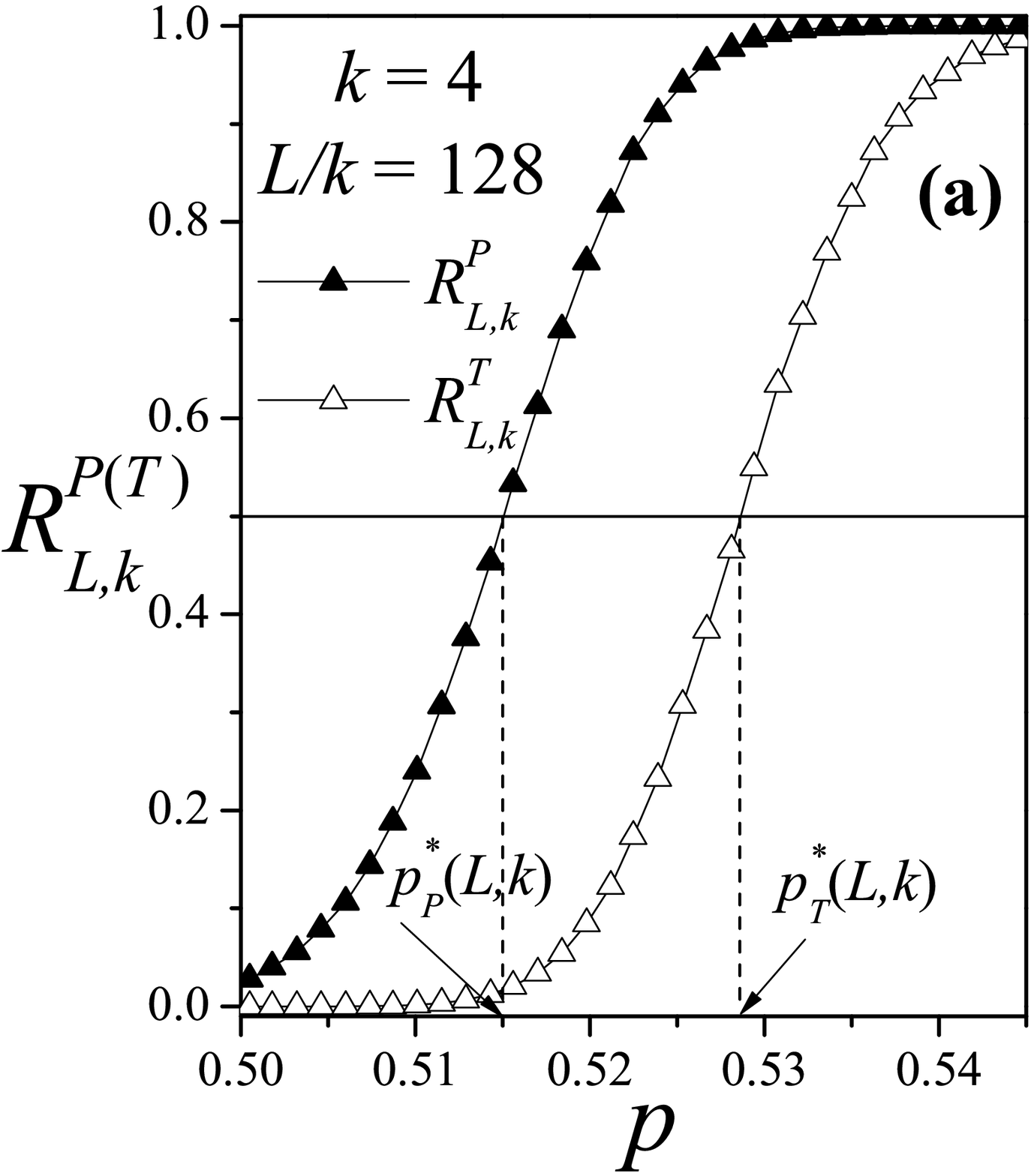}
\includegraphics[width=0.60\columnwidth]{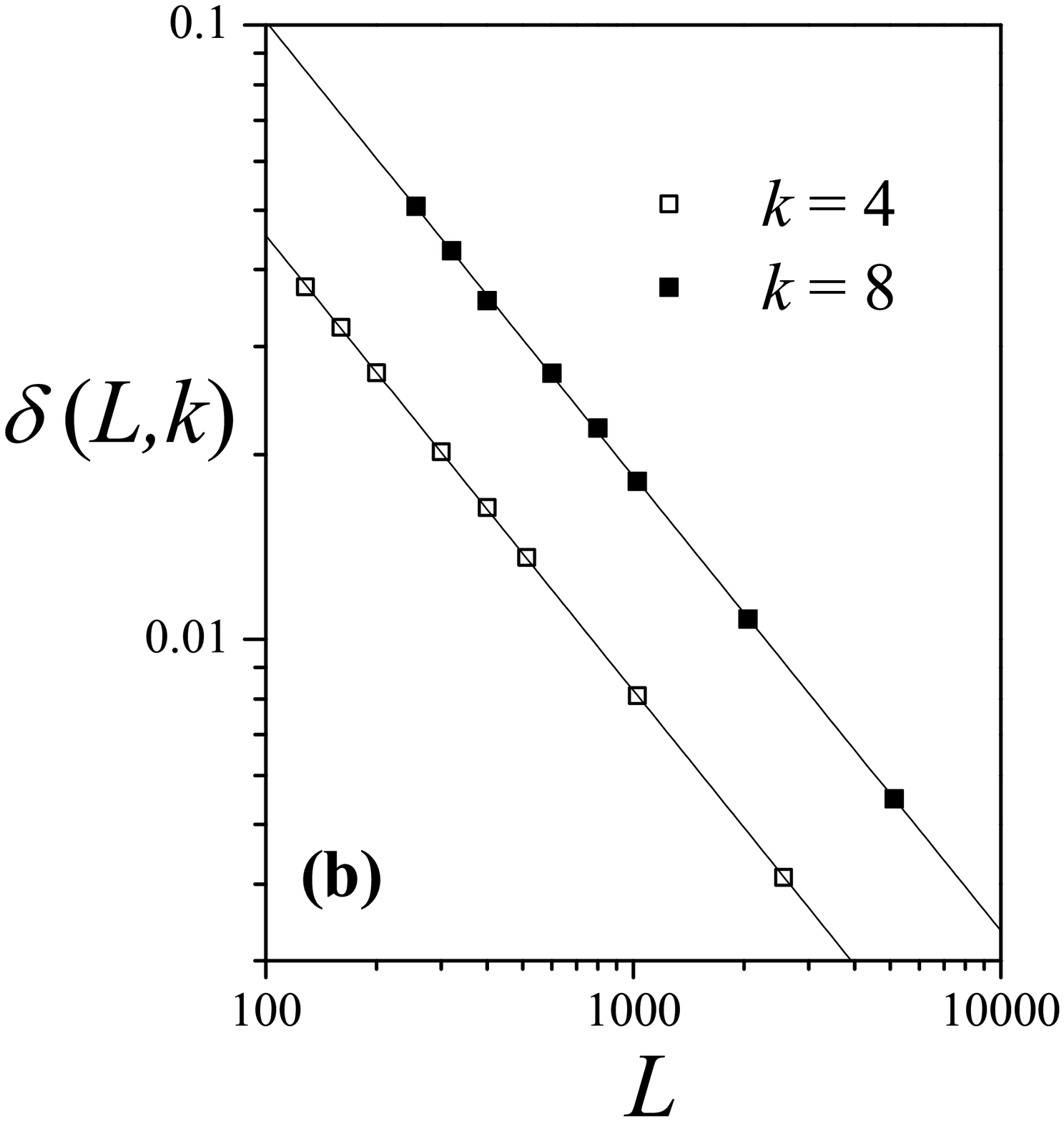}
\caption{(a) $R^P_{L,k}(p)$ (solid triangles) and $R^T_{L,k}(p)$ (open triangles) as a function of the concentration $p$ for $k$ = 4 and $L/k = 128$. The figure illustrates the procedure used to measure the distance $\delta(L,k)$ (see discussion in the text). (b) Log-log plot of $\delta(L,k)$ as a function of the lattice size $L/k$ for two different values of $k$: $k=4$, open squares; and $k=8$, solid squares.}
\label{Figdist}
\end{center}
\end{figure}

An alternative way to visualize the effects described in points $(a)-(b)$ (last paragraph) is presented in Fig. \ref{Figdist}. In part $a)$, the parallel and transversal probabilities are shown for $k=4$ and a fixed value of the lattice size $L/k=128$. In order to measure the separation between the curves in the $p$-space, the distance $\delta(L,k)$ is defined as $\delta(L,k)=p_T^*(L,k)-p_P^*(L,k)$, where $p_{T[P]}^*(L,k)$ is the value of the concentration $p$ for which $R^{T[P]}_{L,k}(p)=0.5$.

$\delta(L,k)$ was calculated for different values of $k$ and $L$. The results are shown in Fig. \ref{Figdist}b) for two $k$ sizes ($k=4$ and $8$) and $L/k$ ranging between 75 and 640 (note the log-log scale in the figure). In all cases, the separation between the parallel and transversal curves diminishes for increasing $L$, being $\delta(L \rightarrow \infty,k)=0$. This result reinforces the arguments given in the discussion of Fig. \ref{FigRTP}. Namely, for an infinite system of aligned $k$-mers on triangular lattices, the properties of the percolating phase are characterized by an unique percolation threshold, regardless of the lattice direction (transversal or parallel to the alignment direction).

Finally, an exhaustive study of critical exponents and universality was carried out. The results of this analysis are presented in the Supplemental Material \cite{SM}. The values obtained for $\nu$, $\gamma$ and $\beta$ verify that, as expected for a system with short-range correlations (it is well known that RSA has very short-range correlations), this problem belongs to the same universality class that the random percolation problem.

\section{Conclusions}\label{conclu}

In this paper, the percolation behavior of aligned rigid rods of length $k$ on 2D triangular (rhombus-shaped) lattices has been investigated by computer simulations and finite-size scaling analysis. The $k$-mers (with $k$ from 2 to 80) were deposited along one of the directions of the lattice, forming a nematic phase. Lattice sizes up to $L/k = 640$ were used.

For each value of $k$, the size of the largest cluster $S_L$ and the probability $R^X_{L,k}(p)$ ($X=P,T,U,I,A$) that a lattice of size $L$ percolates at concentration $p$ were used to obtain the critical point (percolation threshold $p_c(k)$ and intersection point of the probability curves $R^{X^*}_k$) and the critical exponents $\nu$, $\beta$ and $\gamma$ characterizing the phase transition.


The percolation threshold exhibits a monotonic increasing function when it is plotted as a function of the $k$-mer size: $p_c(k)=A+B/(C+\sqrt{k})$, being $A = p_c(k \rightarrow \infty)= 0.582(9)$ the value of the percolation threshold by infinitely long $k$-mers, $B =-0.47(0.21)$ and $C = 5.79(2.18)$. This behavior, which is reported here for the first time, is completely different to that observed for square lattices, where the percolation threshold decreases with $k$ \cite{PRE7,Tara2012}. The present result clearly demonstrates that the structure of the lattice plays a fundamental role in determining the statistics and percolation properties of extended objects. To finish with the analysis of $p_c(k)$, it is important to note that, for all $k$, the percolation threshold of aligned rods is higher than the corresponding one to isotropic $k$-mers \cite{JSTAT7}.


The effect of the anisotropy on the properties of the percolating phase was investigated by following the behavior of $R^P_{L,k}(p)$ (probability of finding a percolating cluster in a parallel direction to the nematic alignment) and $R^T_{L,k}(p)$ (probability of finding a percolating cluster in a transverse direction to the nematic alignment). For finite systems, while in the case of isotropic $k$-mers the vertical and horizontal percolation probabilities are indistinguishable, in the case of aligned $k$-mers the anisotropy of the deposited layer favors the percolation along the direction of the nematic phase. The difference between the parallel and transversal probabilities diminishes for increasing the lattice size $L$, being $R^P_{L,k}(p)=R^T_{L,k}(p)$ in the limit of $L \rightarrow \infty$. In other words, the value of the percolation threshold is the same in both parallel and transversal directions.

The breaking of the orientational symmetry influences also the behavior of the percolation probabilities at the intersection point $R^{X^*}_k$. Thus, $R^{U^*}_k$ and $R^{I^*}_k$ exhibit a nonuniversal critical behavior, varying continuously with changing the $k$-mer size. A similar scenario has been already reported in the case of aligned $k$-mers on square lattices \cite{PRE7,Tara2012} and thermal transitions in the presence of anisotropy \cite{Selke1,Selke2}. These results indicate that the universality of the intersection points $R^{X^*}_k$'s is observed only for isotropic systems. For anisotropic systems, this universality is violated and the value of the crossing point of the percolation probabilities is dependent upon $k$ (and the degree of alignment).


Finally, the improved accuracy in the determination of the critical exponents ($\nu$, $\beta$ and $\gamma$) confirmed that the model belongs to the same universality class as the random percolation, regardless of the size $k$ considered. In addition, the corresponding curves collapse according to the predictions of the scaling theory.

\section{ACKNOWLEDGMENTS}

This work was supported in part by CONICET (Argentina) under project number PIP 112-201101-00615; Universidad Nacional de San Luis (Argentina) under project No. 03-0816; and the National Agency of Scientific and Technological Promotion (Argentina) under project  PICT-2013-1678. The numerical work were done using the BACO parallel cluster (http://cluster\_infap.unsl.edu.ar/wordpress/) located  at Instituto de F\'{\i}sica Aplicada, Universidad Nacional de San Luis - CONICET, San Luis, Argentina.

\end{large}

\end{document}